\documentclass[aps,amsmath,amssymb,amsfonts,twocolumn]{revtex4}
\usepackage{epsfig}
\usepackage{graphicx}
\usepackage{subfigure}
\usepackage{dcolumn}
\usepackage{bm}
\usepackage{amsthm}
\usepackage{enumitem}
\usepackage{slashed}
\usepackage{braket}
\usepackage{amsmath}
\usepackage{mathtools}

\makeatletter
\def\l@subsubsection#1#2{}
\makeatother

\newcommand{\del}{\partial}

\renewcommand{\Re}{\operatorname{Re}}

\newcommand{\calA}{\mathcal{A}}

\newcommand{\calH}{\mathcal{H}}

\newcommand{\calM}{\mathcal{M}}

\begin{document}

\title{Quartic locality of higher-spin gravity in de Sitter and Euclidean anti-de Sitter space}

\author{Yasha Neiman}
\email{yashula@icloud.com}
\affiliation{Okinawa Institute of Science and Technology, 1919-1 Tancha, Onna-son, Okinawa 904-0495, Japan}

\date{\today}

\begin{abstract}
We consider higher-spin gravity in (A)dS\textsubscript{4}, defined as the holographic dual of a free $O(N)$ or $Sp(N)$ vector model. At the quartic level, this theory has been judged non-local at distances greater than the (A)dS radius, due to a mismatch of massless (twist=1) exchange-type terms in its boundary OPE behavior. We review the non-locality argument, and note that it relies on a double-lightcone limit, which requires a Lorentzian boundary. In the Euclidean OPE limit, we demonstrate the absence of massless exchange-type non-localities of any spin, by inspecting a known formula for the bulk exchange diagrams in Euclidean AdS, and constructing upper bounds in which the spin-dependence factorizes from the position-dependence. Our results suggest that higher-spin theory is local (at distances greater than the curvature radius) in spacetimes with Euclidean boundary signature. For Lorentzian bulk, this implies locality in de Sitter space, as opposed to anti-de Sitter.
\end{abstract}

\maketitle
%\tableofcontents
%\newpage

\section{Introduction and summary} \label{sec:intro}

Minimal type-A higher-spin (HS) gravity in $D=4$ spacetime dimensions \cite{Vasiliev:1990en,Vasiliev:1995dn,Vasiliev:1999ba} is the interacting theory of an infinite tower of parity-even massless fields, one for each even spin. It is also the conjectured bulk dual \cite{Klebanov:2002ja} within AdS/CFT \cite{Maldacena:1997re,Gubser:1998bc,Witten:1998qj,Aharony:1999ti} of a particularly simple boundary theory: the free $O(N)$ vector model of $N$ real scalar fields $\varphi^I$ ($I=1,\dots,N$). Remarkably, this holographic duality can be extended from AdS to de Sitter space \cite{Anninos:2011ui}, thus offering a window into 4d quantum gravity with positive cosmological constant. Note that the same bulk theory with different boundary conditions is also dual to the critical (interacting) vector model, while its close relatives are dual to various Chern-Simons-matter theories \cite{Sezgin:2002rt,Sezgin:2003pt,Aharony:2011jz,Giombi:2011kc,Giombi:2012ms}. In this work, we focus on the simplest boundary dual, i.e. the free vector model.

The linearized version of HS theory was first formulated by Fronsdal \cite{Fronsdal:1978rb,Fronsdal:1978vb}. A partial set of the cubic vertices was found by Fradkin and Vasiliev \cite{Fradkin:1986qy,Fradkin:1987ks}. The complete set was found by Sleight and Taronna \cite{Sleight:2016dba}, by matching with the cubic correlators of the boundary theory. To similarly extract the \emph{quartic} vertex from the boundary correlator, 4-point exchange diagrams (with spin-0 external legs) were evaluated in \cite{Bekaert:2015tva}. However, the resulting expressions were formidable, and this ``direct'' line of attack got stalled. Then, independently of the explicit expressions in \cite{Bekaert:2015tva}, Sleight and Taronna noticed \cite{Sleight:2017pcz} a crucial problem at the quartic order: a mismatch in massless (twist=1) exchange-type terms between the boundary correlator and the expected behavior of local bulk diagrams. The upshot is that either the quartic vertex, or the sum over spins in the exchange diagrams, must deviate from local behavior even at distances much larger than the (A)dS radius, which opens the door to uncontrollable field-redefinition freedom. This is in contrast to the initial expectation that HS theory's non-locality should be contained within $\sim 1$ (A)dS radius.

The argument of \cite{Sleight:2017pcz} is holographic. Via the usual UV/IR duality, the question of non-locality at large bulk distances becomes a question of short-distance behavior on the boundary. Accordingly, \cite{Sleight:2017pcz} studies the OPE structure of the boundary correlator and its ingredients. More specifically, as we review in section \ref{sec:Lorentz}, it studies OPE behavior \emph{in two channels at once}. This makes sense in Lorentzian boundary signature, where one point can be lightlike-separated from two others, but not in Euclidean. Alternatively, one can say that \cite{Sleight:2017pcz} is concerned with the bulk diagrams' conformal-block structure, which specifies their behavior at arbitrary complex boundary coordinates. Since complex space allow non-zero null vectors, this is essentially the same as studying the Lorentzian case. 

Generally, the Euclidean OPE limit is more restrictive than the Lorentzian one, since it requires zero (rather than merely lightlike) separation. This means that the Euclidean OPE behavior of bulk diagrams constitutes a less stringent locality requirement. This is widely recognized in the CFT and holography literature, which studies the Lorentzian limit so as to access its more stringent constraints \cite{Fitzpatrick:2012yx,Fitzpatrick:2014vua,Alday:2010zy,Alday:2015ota}. But HS theory can benefit from the opposite: restricting to the Euclidean OPE limit can provide us with a regime where the problematic non-locality of \cite{Sleight:2017pcz} is absent. In section \ref{sec:Euclid}, we'll show that this is indeed the case. Using the explicit formulas for bulk exchanges \cite{Bekaert:2015tva}, we will derive bounds on their Euclidean OPE behavior, and demonstrate the absence of exchange-type non-localities of twist$\,<\,$2 (i.e. ones that would dominate over the local double-trace contributions), at any spin. Note that the formulas of \cite{Bekaert:2015tva} were originally written for Euclidean AdS, but can refer equally to Lorentzian de Sitter, via the analytic continuation of \cite{Maldacena:2002vr,Anninos:2011ui}. The upshot is that HS gravity appears to be unacceptably non-local for Lorentzian boundary (i.e. in Lorentzian AdS), but not for Euclidean boundary (i.e. in Euclidean AdS or Lorentzian de Sitter).

From the bulk point of view, such signature-dependence of non-locality is not so unexpected. For instance, consider a bulk vertex containing an infinite tower of derivatives, which amounts to some finite non-locality radius. This has very different practical meaning in Euclidean (where finite radius describes a compact sphere around a single point) vs. Lorentzian (where it describes a non-compact hyperboloid around a lightcone). In particular, in Lorentzian, one can travel between two distant points via two ``short'' (i.e. near-lightlike) steps, which isn't an option in Euclidean. As a result, the same analytic expression may exhibit sufficiently local behavior in Euclidean, but not in Lorentzian. Similarly, in a Lorentzian bulk, the sign of the cosmological constant will also affect the notion of locality, by changing the asymptotic behavior of lightcones. In the end, it seems that the main variable is the signature of the boundary, as it controls the possible types of large-distance bulk separations. 

We stress that for all bulk/boundary signatures, the bulk vertices should still be given by the same analytic expressions. Thus, our Euclidean result rules out a simple massless ``$\Box^{-1}$'' non-locality \emph{for all signatures}, and implies that the (Lorentzian) non-locality predicted by \cite{Sleight:2017pcz} must take a more subtle form.

Alongside the main results, this paper contains two Appendices. Appendix \ref{app:spin_0} briefly discusses spin-0 exchanges, since the formulas of \cite{Bekaert:2015tva} only apply to nonzero spins. Appendix \ref{app:single} addresses a tension between our Euclidean results and a certain expectation in the HS community, namely that bulk exchange diagrams should be qualitatively well-described by their single-trace pieces \emph{even outside} the relevant channel's OPE limit.

\section{General expectations from a local bulk theory} \label{sec:general}

We work in AdS\textsubscript{4} with unit curvature radius, in Poincare coordinates $(z,\mathbf{x})$, where the boundary coordinates $\mathbf{x}$ may be Lorentzian (section \ref{sec:Lorentz}) or Euclidean (section \ref{sec:Euclid}). We denote boundary distances as $r_{ij} = |\mathbf{r}_{ij}|$, where $\mathbf{r}_{ij} = \mathbf{x}_j-\mathbf{x}_i$. We will be interested in the connected 4-point functions $\langle J_1 J_2 J_3 J_4 \rangle$ of the single-trace scalar operator $J(\mathbf{x}) = \frac{c}{2}\varphi_I(\mathbf{x})\varphi^I(\mathbf{x})$ (with normalization coefficient $c=\frac{4}{\sqrt{N}}$). The theory's other single-trace operators are the spin-$s$ conserved currents, having conformal weights $\Delta=s+1$, i.e. twist=1.  As usual, the 4-point function can be expanded in conformal blocks in any exchange channel $(ij|kl)$: 
\begin{align}
    \langle J_1 J_2 J_3 J_4 \rangle = \calH_{ij|kl}^{\text{single}} + \calH_{ij|kl}^{\text{double}} \ . \label{eq:OPE}
\end{align}
In the OPE limit $r_{ij}r_{kl}\to 0$, the double-trace piece $\calH_{ij|kl}^{\text{double}}$ is an analytic function of the coordinates $\mathbf{x}_i$, while the single-trace piece $\calH_{ij|kl}^{\text{single}}$ is an analytic function multiplied by the divergent factor $1/(r_{ij}r_{kl})$. Note that a general double-trace object may also include an anomalous-dimension term, which is an analytic function multiplied by $\ln(r_{ij}r_{kl})$. Such a term is absent in $\calH_{ij|kl}^{\text{double}}$, due to the CFT being free. Now, the same 4-point function \eqref{eq:OPE} can be written as a sum of bulk Witten diagrams:
\begin{align}
  \langle J_1 J_2 J_3 J_4 \rangle = \calA_{12|34} + \calA_{13|24} + \calA_{14|23} + \calA_{\text{contact}} \ . \label{eq:Witten_diagrams} 
\end{align}
Here, the exchange diagrams $\calA_{ij|kl}$ are built from the known propagators \cite{Bekaert:2014cea} and cubic vertices \cite{Bekaert:2015tva}, with the quartic-vertex diagram $\calA_{\text{contact}}$ then \emph{defined} by \eqref{eq:Witten_diagrams}. In a local bulk theory, we have some expectations of the bulk diagrams' behavior in an OPE limit $r_{ij}r_{kl}\to 0$. In the Lorentzian (or complex) case, these are encoded as an expected conformal-block structure of the bulk diagrams \cite{Heemskerk:2009pn,Heemskerk:2010ty,El-Showk:2011yvt}. Specifically, the direct-channel exchange $\calA_{ij|kl}$ should be given by the single-trace piece $\calH_{ij|kl}^{\text{single}}$ from \eqref{eq:OPE} plus some double-trace ``dressing'', while the cross-channel exchanges $\calA_{ik|jl},\calA_{il|jk}$ and the quartic-vertex diagram $\calA_{\text{contact}}$ should be double-trace. This means that $\calA_{ij|kl}$ should diverge as a power law $\sim 1/(r_{ij}r_{kl})$, while $\calA_{ik|jl},\calA_{il|jk},\calA_{\text{contact}}$ should diverge at most as $\sim\ln(r_{ij}r_{kl})$. In bulk terms, this difference is due to ``short'' vs. ``long'' boundary-bulk propagators, and is insensitive to non-localities confined within $\sim 1$ AdS radius. Therefore, a \emph{power-law divergence in the wrong place} signals an unacceptable level of non-locality. Such ``exchange-type non-localities'' are the topic of \cite{Sleight:2017pcz} and of the present work. Note that:
\begin{enumerate}
    \item Eqs. \eqref{eq:OPE}-\eqref{eq:Witten_diagrams} imply that if the bulk exchanges have the correct local structure, then so does $\calA_{\text{contact}}$.
    \item Since the cubic vertices are local, the exchanges \emph{of each individual field} always have the correct local structure. The question is whether this survives the infinite sum over spins.
\end{enumerate}

\section{The Lorentzian non-locality argument} \label{sec:Lorentz}

With these generalities in place, let us compute the actual 4-point function $\langle J_1 J_2 J_3 J_4 \rangle$ in the free vector model. It is given by a sum of three 1-loop Feynman diagrams:
\begin{align}
    \begin{split}
        \langle J_1 J_2 J_3 J_4 \rangle &= \calM_{1234} + \calM_{1243} + \calM_{1324} \ ; \\
        \calM_{ijkl} &= Nc^4 G_{ij}G_{jk}G_{kl}G_{li} \ ,
    \end{split} \label{eq:correlator}
\end{align}
where $G_{ij} = 1/(4\pi r_{ij})$ is the propagator for the fundamental boundary fields $\varphi^I(\mathbf{x})$. We see that e.g. the boundary diagram $\calM_{1234}$ contains a product of $1/(r_{12}r_{34})$ and $1/(r_{23}r_{41})$, i.e. of power-law divergences belonging to \emph{two} OPE limits. In Lorentzian (or complex) coordinates, both divergences can be realized together by taking the ``double-lightcone limit'' \cite{Alday:2010zy,Alday:2015ota}, where e.g. $\mathbf{x}_2$ is lightlike-separated from both $\mathbf{x}_1$ and $\mathbf{x}_3$. In this limit, the doubly-divergent diagram $\calM_{1234}$ dominates the correlator. But which of the bulk diagrams in \eqref{eq:Witten_diagrams} carries this doubly-divergent term? The observation of \cite{Sleight:2017pcz} is that there's no apparent answer to this question that is consistent with bulk locality: as we saw, in a local theory, each diagram in \eqref{eq:Witten_diagrams} should have a power-law divergence in at most one OPE channel, never in two at once. 

Note that, though the text of \cite{Sleight:2017pcz} contains the above form of the argument (section 3.1 therein), its main emphasis is instead on the single/double-trace decomposition \eqref{eq:OPE} of the correlator \eqref{eq:correlator}:
\begin{align}
   \calH_{12|34}^{\text{single}} = \calM_{1234} + \calM_{1243} \ ; \quad \calH_{12|34}^{\text{double}} = \calM_{1324} \ , \label{eq:OPE_diagrams}
\end{align}
which (together with its permutations) implies:
\begin{align}
    \calH_{12|34}^{\text{single}} + \calH_{13|24}^{\text{single}} + \calH_{14|23}^{\text{single}} = 2\langle J_1 J_2 J_3 J_4 \rangle \ . \label{eq:factor_2}
\end{align}
It's easy to see that \eqref{eq:OPE_diagrams} and the factor of 2 in \eqref{eq:factor_2} are closely related to the doubly-divergent structure of $\calM_{ijkl}$ discussed above. Note that, while \eqref{eq:factor_2} holds irrespective of OPE limits, any simple relationship between $\calH_{ij|kl}^{\text{single}}$ and the bulk exchanges $\calA_{ij|kl}$ \emph{does} require such limits. 

\section{Locality for Euclidean boundary} \label{sec:Euclid}

Having reviewed the non-locality problem in Lorentzian/complex coordinates, let's now consider the weaker requirements for locality in Euclidean. Here, the only way to approach an OPE limit (henceforth, $r_{12}r_{34}\to 0$) is to take $\mathbf{r}_{12}$ or $\mathbf{r}_{34}$ vanishing, rather than lightlike. For concreteness, we can use conformal symmetry to fix:
\begin{align}
    \mathbf{x}_1 = 0 \ ; \quad \mathbf{x}_3 \equiv \mathbf{n} \text{ (a unit vector)} \ ; \quad \mathbf{x_4} \to \infty \ , \label{eq:point_fixing}
\end{align}
with $\mathbf{x}_2\equiv\mathbf{r}$ left arbitrary. Our OPE limit is now $\mathbf{r} = \mathbf{r}_{12}\to 0$. Any function of the $\mathbf{x}_i$ can be expanded in multipoles as $\sum_{\ell=0}^\infty r^\ell f_\ell(r^2)P_\ell\big(\frac{\mathbf{r\cdot n}}{r}\big)$, where $P_\ell$ are the Legendre polynomials, and $f_\ell(r^2)$ are some coefficients. Analytic functions of the coordinates correspond to analytic $f_\ell(r^2)$. As in the previous section, the correlator and the individual bulk exchanges have the structure:
\begin{align}
     &\langle J_1 J_2 J_3 J_4 \rangle = \frac{\text{analytic}}{r} + \text{analytic} \ ; \label{eq:correlator_terms} \\
     &\calA_{12|34}^{(s)} = \frac{\text{analytic}}{r} + \text{analytic}\times\ln r + \text{analytic} \ ; \label{eq:direct_terms} \\
     &\calA_{13|24}^{(s)},\calA_{14|23}^{(s)} = \text{analytic}\times\ln r + \text{analytic} \ , \label{eq:cross_terms}
\end{align}
where the $\sim 1/r$ terms in \eqref{eq:direct_terms}, summed over the exchanged spin $s$, reproduce the $\sim 1/r$ term in \eqref{eq:correlator_terms}. Our locality criterion is then that the sum over spins of the analytic and $\sim\ln r$ terms in \eqref{eq:direct_terms}-\eqref{eq:cross_terms} doesn't produce any additional power-law divergence at $\mathbf{r}\to 0$. Specifically, we want to show that these terms, even when summed over $s$, have their multipole coefficients bounded as $f_\ell(r^2) = O(\ln r)$ for all $\ell$. In particular, this will rule out singularities of the type $f_\ell(r^2)\sim 1/r$, which correspond to a massless spin-$\ell$ exchange-type non-locality. Note that we leave open the question of singularities smaller than $\sim\ln r$. A more complete locality analysis should address these as well.

Now, for the direct-channel exchange \eqref{eq:direct_terms}, the desired behavior is easy to demonstrate by a general argument. The analytic and $\sim \ln r$ terms in \eqref{eq:direct_terms} can be expanded in monomials $f_\ell(r^2)\sim r^{2n}$ and $f_\ell(r^2)\sim r^{2n}\ln r$, respectively. On the other hand, for each spin $s$, the exchange-type singularity $\sim r^{s-1}P_s\big(\frac{\mathbf{r\cdot n}}{r}\big)$ is the dominant term in the small-$\mathbf{r}$ expansion of $\calA_{12|34}^{(s)}$. This means that, in the sum over $s$, the coefficient of each monomial $f_\ell(r^2)\sim r^{2n},r^{2n}\ln r$ gets contributions only from a finite range $0\leq s\leq\ell+2n$, and therefore remains finite. Thus, the sum over $s$ cannot be asymptotically larger than the leading term $f_\ell(r^2)\sim \ln r$.

\subsection{The bound on the cross-channel exchanges (ruling out exchange-type non-localities of spin $\ell=0$)} \label{sec:Euclid:ell_0}

For the cross-channel exchanges \eqref{eq:cross_terms}, the proof is more involved, since each term in the multipole expansion $f_\ell(r^2)$ now receives contributions from all exchanged spins. In this case, our approach to demonstrating $f_\ell(r^2) = O(\ln r)$ is to inspect the explicit formulas \cite{Bekaert:2015tva} for the cross-channel exchanges $\calA_{13|24},\calA_{14|23}$, and demonstrate that \emph{the traceless, $\mathbf{r}$-transverse projection of their $\ell$'th tensor derivative $(\del/\del\mathbf{r})^{\otimes\ell}$} is bounded (in its Euclidean norm, or componentwise in a Cartesian basis) as $O(\ln r)$. Here, the purpose of the traceless-transverse projection is to remove the power-law-divergent derivatives of the allowed $\sim \ln r$ factors. For clarity, we begin with $\ell=0$, i.e. proving that $\calA_{13|24},\calA_{14|23}$ themselves are bounded as $O(\ln r)$.

The spin-$s$ bulk exchanges $\calA^{(s)}_{ij|kl}$ have been computed \cite{Bekaert:2015tva} using the split representation of bulk-bulk propagators \cite{Leonhardt:2003qu,Leonhardt:2003sn,Costa:2014kfa,Bekaert:2014cea}. To keep the formulas more symmetric, we revert from \eqref{eq:point_fixing} to arbitrary boundary points $\mathbf{x}_i$, with our OPE limit remaining $\mathbf{r}_{12}\to 0$. Combining eqs. (3.8)-(3.11) of \cite{Bekaert:2015tva} and simplifying the $\Gamma$-function expressions therein, we get the following formula for e.g. $\calA_{13|24}^{(s)}$:
\begin{widetext}
    \begin{align}
        \begin{split}
            \calA_{13|24}^{(s)} ={}& \frac{16 s^2}{N\pi^3(2s)!} \int_{-\infty}^\infty d\nu\,\frac{\nu\tanh(\pi\nu)}{\nu^2 + \left(s - \frac{1}{2}\right)^2} \left|\Gamma\!\left(\frac{1}{2} + s + i\nu\right)\right|^2 \frac{1}{r_{13}^{\frac{1}{2}-i\nu} r_{24}^{\frac{1}{2}+i\nu}} \int d^3\mathbf{x}_0\,\frac{P_s(\mathbf{v}_{0;13}\cdot\mathbf{v}_{0;24})}{(r_{01}r_{03})^{\frac{3}{2}+i\nu} (r_{02}r_{04})^{\frac{3}{2}-i\nu}} \ ,
        \end{split} \label{eq:exchange}
    \end{align}
\end{widetext}
where the point $\mathbf{x}_0$ is integrated over the boundary, $\mathbf{v}_{0;ij}$ are unit vectors given by:
\begin{align}
    \mathbf{v}_{0;ij} =  \frac{1}{r_{ij}}\left(\frac{r_{0j}}{r_{0i}}\mathbf{r}_{0i} - \frac{r_{0i}}{r_{0j}}\mathbf{r}_{0j} \right) \ ,
\end{align}
and $P_s$ is again the Legendre polynomial:
\begin{align}
    P_s(\mathbf{v}_{0;13}\cdot\mathbf{v}_{0;24}) = \frac{1}{2^s}\binom{2s}{s}\,\mathbf{v}_{0;13}^{\otimes s} \cdot\big(\mathbf{v}_{0;24}^{\otimes s} - \text{traces}\big) \ . \label{eq:Legendre}
\end{align}
To get an upper bound on the absolute value of \eqref{eq:exchange}, we replace all the integrands with their absolute values. This means removing all the imaginary $i\nu$ exponents, and replacing $P_s(\dots)$ with its absolute value, which is $\leq 1$. We can also replace $\nu\tanh(\pi\nu)$ with $|\nu|$. We end up with a factorized bound:
\begin{align}
   \left|\calA^{(s)}_{13|24}\right| \leq \frac{32}{N\pi^3\sqrt{r_{13}r_{24}}}\times C_s\times F(\mathbf{x}_1,\mathbf{x}_2,\mathbf{x}_3,\mathbf{x}_4) \ , \label{eq:bound}
\end{align}
where $C_s$ is a $d\nu$ integral that depends only on the spin $s$, while $F$ is a $d^3\mathbf{x}_0$ integral that depends only on the positions $\mathbf{x}_1,\mathbf{x}_2,\mathbf{x}_3,\mathbf{x}_4$:
\begin{align}
  &C_s = \frac{s^2}{(2s)!} \int_0^\infty \frac{\nu\,d\nu}{\nu^2 + \left(s - \frac{1}{2}\right)^2} \left|\Gamma\!\left(\frac{1}{2} + s + i\nu\right)\right|^2 ; \label{eq:C} \\
  &F(\mathbf{x}_1,\mathbf{x}_2,\mathbf{x}_3,\mathbf{x}_4) = \int \frac{d^3\mathbf{x}_0}{(r_{01}r_{02}r_{03}r_{04})^{3/2}} \ . \label{eq:F}
\end{align}
Let us now estimate the spin-dependent integral \eqref{eq:C}. Since our concern is the infinite sum over $s$, we focus on $s\gg 1$. We can then replace $\big(s - \frac{1}{2}\big)$ with $s$, and use Stirling's formula for the factorial and $\Gamma$-function as:
\begin{align}
  \frac{1}{(2s)!}\left|\Gamma\!\left(\frac{1}{2}+s+i\nu\right)\right|^2 \approx \sqrt{\frac{\pi}{s}}\left(\frac{e^{-2t\arctan t}(1+t^2)}{4}\right)^s \ , 
\end{align}
where we denoted $t\equiv \nu/s$. We notice that the numerator $e^{-2t\arctan t}(1+t^2)$ is bounded as $\leq 1$, having its global maximum at $t=0$. We can therefore remove it from the $(\dots)^s$ exponential. Plugging everything back into \eqref{eq:C}, we get:
\begin{align}
  C_s \leq \frac{\sqrt{\pi}\,s^{3/2}}{4^s}\,\int_0^\infty dt\,t\,e^{-2t\arctan t} \ . \label{eq:Stirling_bound}
\end{align}
The integral converges, and we conclude (neglecting the $s^{3/2}$ power law in favor of the $4^{-s}$ exponential):
\begin{align}
    C_s \lesssim 4^{-s} \ . \label{eq:C_scaling}
\end{align}
In fact, evaluating \eqref{eq:C} numerically, one finds $C_s\sim 4^{-s}$. The upshot is that the sum $\sum_s C_s$ rapidly converges. 

Next, consider the boundary integral \eqref{eq:F}. This is clearly finite for distinct points $\mathbf{x}_1,\mathbf{x}_2,\mathbf{x}_3,\mathbf{x}_4$. In the limit $\mathbf{r}_{12}\to 0$, the integral is dominated by $\mathbf{x}_0$ close to $\mathbf{x}_1,\mathbf{x}_2$, and can be estimated by counting powers of the small vectors $\mathbf{r}_{01},\mathbf{r}_{02},\mathbf{r}_{12}$ in the integrand. In the case of \eqref{eq:F}, the lowest power is $-3$, so the integral diverges at most logarithmically. Combined with the convergence of $\sum_s C_s$, this implies that the sum over $s$ of the cross-channel exchanges \eqref{eq:exchange},\eqref{eq:bound} is indeed bounded as $\calA_{13|24} = O(\ln r_{12})$.

\subsection{Generalizing the bound to higher multipoles (ruling out exchange-type non-localities of spin $\ell>0$)}

We are now ready to address the higher multipoles $\ell>0$ of the cross-channel exchanges. Upon taking the traceless, $\mathbf{r}_{12}$-transverse projection of the $\ell$'th-order tensor derivative w.r.t. $\mathbf{x}_{2}$, we employ the same strategy as above: we seek a bound that factorizes into a $s$-dependent $d\nu$ integral that scales as \eqref{eq:C_scaling}, and a position-dependent $d^3\mathbf{x}_0$ integral that diverges at most logarithmically. To avoid cumbersome formulas, we'll use our $\ell=0$ analysis above as a reference point, and list the extra steps/complications that arise for $\ell>0$:
\begin{enumerate}
    \item The $\del/\del\mathbf{x}_2$ derivatives must be applied to \eqref{eq:exchange} before any absolute values or bounds. Since \eqref{eq:exchange} has several $\mathbf{x}_2$-dependent factors, the derivatives yield a sum of terms, whose number grows with $\ell$ but not with $s$. Our bound procedure can be applied to each of these terms separately.
    \item When hitting the powers of $r_{0i},r_{ij}$ in \eqref{eq:exchange}, the derivatives produce a polynomial in $\nu$, of order $O(\ell)$. When constructing the factorized bound, this can be tacked onto the $s$-dependent $d\nu$ integral, without affecting its dominant scaling \eqref{eq:C_scaling}.
    \item The derivatives can also hit the Legendre polynomial $P_s$ in \eqref{eq:exchange}, replacing it by its derivatives of order $k\leq\ell$. For $k>s$, these vanish. For $k\leq s$, they (just like $P_s$ itself) are bounded by their value at the endpoints, as:
    \begin{align}
        \left|\frac{d^kP_s(t)}{dt^k}\right| \leq \left.\frac{d^kP_s(t)}{dt^k}\right|_{t=1} = \frac{(s+k)!}{2^k k! (s-k)!} \ .
    \end{align}
   This is a polynomial in $s$ of order $2k$, which again can be tacked onto the $s$-dependent $d\nu$ integral without affecting its dominant behavior \eqref{eq:C_scaling}.
   \item Some of the derivatives lower the power of the small vectors $\mathbf{r}_{01},\mathbf{r}_{02},\mathbf{r}_{12}$ in the position-dependent $d^3\mathbf{x}_0$ integral, threatening a power-law divergence. However, these same derivatives produce either a tensor index that points \emph{along} one of these vectors, or a factor of the metric. This is where the traceless-transverse projection becomes important. First, it kills terms with factors of the metric or indices pointing along $\mathbf{r}_{12}$. Second, it requires indices along $\mathbf{r}_{01},\mathbf{r}_{02}$ to be accompanied by contractions of $\mathbf{r}_{01},\mathbf{r}_{02}$ with one of $\mathbf{r}_{03},\mathbf{r}_{04},\mathbf{r}_{13},\mathbf{r}_{24}$, which serves to raise the power of the small vectors $\mathbf{r}_{01},\mathbf{r}_{02},\mathbf{r}_{12}$ back up to $-3$, as needed for a $O(\ln r_{12})$ result.
\end{enumerate}
This concludes our argument that, for Euclidean boundary, the multipole coefficients $f_\ell(r_{12}^2)$ in the cross-channel exchanges remain bounded as $O(\ln r_{12})$ even after the sum over spins $s$, as expected for non-locality confined within $\sim 1$ curvature radius.

\begin{acknowledgments}
I am grateful to Ofer Aharony, Shai Chester, Olga Gelfond, Slava Lysov, Dmitry Ponomarev, Evgeny Skvortsov, Mirian Tsulaia and Mikhail Vasiliev for discussions. Some of these took place at the workshop ``Higher spin gravity and its applications'' at APCTP, Pohang. I especially thank Charlotte Sleight and Massimo Taronna for their patience in helping me understand their work, and for their help in correctly assembling eq. \eqref{eq:exchange}. I am supported by the Quantum Gravity Unit of the Okinawa Institute of Science and Technology Graduate University (OIST).
\end{acknowledgments}

\appendix
\section{Spin-0 exchanges} \label{app:spin_0}

For $s=0$, the general bulk exchange formula \eqref{eq:exchange} doesn't apply. Specifically, \eqref{eq:exchange} in this case evaluates to 0, due to the cubic coupling $g_s\sim\frac{1}{\Gamma(D-4+s)}$ vanishing at $D=4,s=0$. In our case of a free vector model on the boundary (as opposed to the critical vector model), this behavior is incorrect. It arises from eq. (3.3) of \cite{Bekaert:2015tva}, which produces the scalar bulk-bulk propagator with vanishing weight-1 boundary data, whereas we need the one with vanishing weight-2 boundary data. With the correct propagator, the vanishing couplings cancel against divergent bulk integrals to give a finite result, which can be evaluated directly (i.e. without the split-representation technique of \cite{Bekaert:2015tva}). Dimensional regularization leads to a simple prescription: we must integrate the cubic vertices over the boundary rather than bulk, replace all boundary-bulk and bulk-bulk propagators by the 2-point function $\langle J_i J_j\rangle = 1/(2\pi^2 r_{ij}^2)$, and assign an effective coupling $\tilde g_0 = \frac{8}{\sqrt{N}}$ to each vertex. This leads to the formula:
\begin{align}
    \calA_{ij|kl}^{(0)} = \frac{2}{\pi^{10}N}\int\frac{d^3\mathbf{x}_0\,d^3\mathbf{x}'_0}{r_{0i}^2 r_{0j}^2 r_{00'}^2 r_{k0'}^2 r_{l0'}^2} \ , \label{eq:scalar_exchange}
\end{align}
where $\mathbf{x}_0,\mathbf{x}'_0$ are the positions of the two vertices. 

While slightly unusual, eq. \eqref{eq:scalar_exchange} has the standard OPE properties of a bulk exchange diagram. For instance, in the Euclidean OPE limit $\mathbf{r}_{12}\to 0$, the direct-channel exchange $\calA_{12|34}^{(0)}$ diverges as $\sim 1/r_{12}$ (with the integral dominated by $\mathbf{x}_0$ close to $\mathbf{x}_1,\mathbf{x}_2$), while the cross-channel exchanges $\calA_{13|24}^{(0)},\calA_{14|23}^{(0)}$ diverge as $\sim \ln r_{12}$ (with the integral dominated by both $\mathbf{x}_0$ and $\mathbf{x}'_0$ close to $\mathbf{x}_1,\mathbf{x}_2$). 

The distinct formula \eqref{eq:scalar_exchange} for $s=0$ doesn't matter much for the Euclidean locality argument of section \ref{sec:Euclid}. The OPE behavior of individual-spin bulk exchanges is in any case well-understood, and our main focus in section \ref{sec:Euclid} was on taming the infinite sum over $s$.
\\
\section{More on the difference between $\calA_{ij|kl}$ and $\calH_{ij|kl}^{\text{single}}$} \label{app:single}

Our results in section \ref{sec:Euclid} imply that in the Euclidean OPE limit $\mathbf{r}_{12}\to 0$, the cross-channel bulk exchanges $\calA_{13|24}+\calA_{14|23}$ behave very differently from their single-trace parts $\calH_{13|24}^{\text{single}}+\calH_{14|23}^{\text{single}}$. Indeed, we showed that the former behave as $O(\ln r_{12})$, whereas the latter yield a $\sim 1/r_{12}$ exchange-type singularity, as one can see by rearranging \eqref{eq:OPE_diagrams}-\eqref{eq:factor_2} as:
\begin{align}
    \calH_{13|24}^{\text{single}} + \calH_{14|23}^{\text{single}} = \calH_{12|34}^{\text{single}} + 2\calH_{12|34}^{\text{double}} \ . \label{eq:HS_crossing}
\end{align}
This starkly different behavior goes against a widespread expectation in the HS community, namely that $\calA_{ij|kl}$ and $\calH_{ij|kl}^{\text{single}}$ should be similar ``up to local corrections'', not only in the same-channel OPE limit, but always. In this Appendix, we set out to defuse this false expectation, by listing specific differences between the two objects, and offering some intuitions for why these differences occur. We work in Euclidean AdS.

To put $\calA_{ij|kl}$ and $\calH_{ij|kl}^{\text{single}}$ on a similar footing, we express both in terms of bulk diagrams. For $\calH_{ij|kl}^{\text{single}}$, this is accomplished via geodesic Witten diagrams \cite{Hijano:2015zsa,Dyer:2017zef}. In this approach, the contribution to $\langle J_1 J_2 J_3 J_4\rangle$ of a particular conformal block in the $(ij|kl)$ channel is computed in terms of two bulk geodesics exchanging a bulk field, with one geodesic stretching between the boundary points $\mathbf{x}_i,\mathbf{x}_j$, and the other between $\mathbf{x}_k,\mathbf{x}_l$. Substituting the HS gravity multiplet as the exchanged field (and summing over spin), we get precisely the contribution $\calH_{ij|kl}^{\text{single}} = \sum_s \calH_{ij|kl}^{\text{single}(s)}$ of the single-trace conformal blocks. This picture was explored in detail in \cite{David:2020fea,Lysov:2022zlw}, where it was noticed that the resulting geodesics (with their specific couplings to the HS multiplet) can be thought of as the worldlines of linearized Didenko-Vasiliev ``black holes'' \cite{Didenko:2008va,Didenko:2009td}. Strictly speaking, the topic of \cite{David:2020fea,Lysov:2022zlw} was not the OPE of $J(\mathbf{x}_i)J(\mathbf{x}_j)$, but the bilocal single-trace operator $\varphi_I(\mathbf{x}_i)\varphi^I(\mathbf{x}_j)$. However, from the boundary Feynman diagrams \eqref{eq:correlator}-\eqref{eq:OPE_diagrams}, one can see that, up to normalization, the single-trace part of $J(\mathbf{x}_i)J(\mathbf{x}_j)$ is just $\varphi_I(\mathbf{x}_i)\varphi^I(\mathbf{x}_j)$ multiplied by a boundary propagator $G_{ij} = 1/(4\pi r_{ij})$. With these adjustments, we can read off from eqs. (103),(146) of \cite{David:2020fea} the single-trace spin-$s$ piece $\calH_{ij|kl}^{\text{single}(s)}$ in e.g. the $(13|24)$ channel as:
\begin{widetext}
 \begin{align}
    \begin{split}  
       \calH_{13|24}^{\text{single}(s)} ={}& \frac{16}{\pi^5N(r_{13}r_{24})^2}\int_{\sinh\chi}^\infty\frac{dR}{\sqrt{(R^2 - \sinh^2\chi)(R^2 + \sin^2\theta)}} \\
          &\qquad\times \left\{\begin{array}{cc}
              \frac{1}{2} & \quad s=0 \\[0.3em]
              \displaystyle \frac{1}{(R^2+1)^s}\Re \left(\cosh\chi\cos\theta + \frac{i}{R}\sqrt{(R^2 - \sinh^2\chi)(R^2 + \sin^2\theta)} \right)^s & \quad s>0
          \end{array} \right. \ .
    \end{split} \label{eq:geodesic} 
 \end{align}
\end{widetext}
The integral in \eqref{eq:geodesic} is over one of the two bulk geodesics, where $R$ is the hyperbolic sine of the distance between the integration point and the other geodesic. The parameters $\chi,\theta$ are respectively the distance and angle between the two geodesics at their point of closest approach. These are determined by the boundary points $\mathbf{x}_i$ via eq. (101) of \cite{David:2020fea}:
\begin{align}
 \begin{split}
   \cosh\chi &= \frac{r_{14}r_{23} + r_{12}r_{34}}{r_{13}r_{24}} \ ; \\ 
   \cos\theta &= \frac{r_{14}r_{23} - r_{12}r_{34}}{r_{13}r_{24}} \ .
 \end{split} \label{eq:chi_theta}
\end{align}
We are now ready to list some of the key differences between $\calH_{13|24}^{\text{single}}$ and $\calA_{13|24}$. 

\paragraph{Divergences outside the OPE limit (for any fixed spin).}
The spin-$s$ contributions to $\calA_{13|24}$ diverge only when some of the boundary points $\mathbf{x}_i$ coincide (recall that we are working in Euclidean). Thanks to our bound \eqref{eq:bound}-\eqref{eq:F}, we see that this remains true after summing over spins. In contrast, the spin-$s$ contributions to $\calH_{13|24}^{\text{single}}$ diverge when the two bulk geodesics intersect (i.e. when $\chi=0$), which happens whenever the four boundary points $\mathbf{x}_i$ lie on the same circle, in a cyclic ordering where $\mathbf{x}_1$ and $\mathbf{x}_3$ are non-adjacent. These spurious divergences cancel when summed over spin \cite{David:2020fea}, in what amounts to a HS extension of the well-known cancellation of the electric and gravitational forces between BPS objects in supergravity.

\paragraph{Degree of divergence in the cross-channel OPE limit (for any fixed spin).}
Now, consider the OPE limit $\mathbf{r}_{12}\to 0$. As we saw in section \ref{sec:Euclid:ell_0}, the bulk exchange $\calA_{13|24}$ in this limit (both for individual spins and for their sum) diverges at most as $\sim \ln(r_{12}r_{34})$. In the geodesic diagram \eqref{eq:geodesic}-\eqref{eq:chi_theta}, the limit corresponds to $\chi$ and $\theta$ both vanishing, as:
\begin{align}
    \chi^2 + \theta^2 \approx \frac{4r_{12}r_{34}}{r_{13}r_{24}} \ ,
\end{align}
while the ratio $\theta/\chi$ interpolates between the case of parallel geodesics ($\theta=0$) and intersecting geodesics ($\chi=0$). In this limit, the integral \eqref{eq:geodesic} is dominated by $R\ll 1$, and its spin-dependent last line can be approximated as $\approx 1$. We then evaluate the integral in the first line, which gives:
\begin{align}
  \calH_{13|24}^{\text{single}(s)} \approx{}& \frac{8}{\pi^5N(r_{13}r_{24})^{3/2}\sqrt{r_{12}r_{34}}}\,K\!\left(\sqrt{\frac{\theta^2}{\chi^2+\theta^2}}\right) \nonumber \\
    & \times \left\{\begin{array}{cc}
        \frac{1}{2} & \quad s=0 \\[0.3em]
        1 & \quad s>0
    \end{array} \right. \ , \label{eq:H_limit}
\end{align}
where $K$ is the complete elliptic integral of the first kind. We see that $\calH_{13|24}^{\text{single}(s)}$ diverges as $\sim 1/\sqrt{r_{12}}$, i.e. stronger than $\calA^{(s)}_{13|24}$. From the bulk point of view, this is easy to understand. In the standard exchange diagram $\calA^{(s)}_{13|24}$, nothing forces the two cubic vertices to be very close to each other. In contrast, in the geodesic diagram $\calH_{13|24}^{\text{single}(s)}$, the vertices are forced to lie on the geodesics, which come very close in the $\mathbf{r}_{12}\to 0$ limit.

Note that for $\theta=0$, the divergence of $\calH_{13|24}^{\text{single}(s)}$ takes the form $\sim 1/\chi$, which is just the (spin-$s$) Coulomb potential between the two geodesics at their point of closest approach. 

Note also that the $\sim 1/\sqrt{r_{12}}$ divergence of $\calH_{13|24}^{\text{single}(s)}$, while stronger than that of $\calA^{(s)}_{13|24}$, is still weaker than the $\sim 1/r_{12}$ divergence of the (summed over spin) combination $\calH_{13|24}^{\text{single}}+\calH_{14|23}^{\text{single}}$ in \eqref{eq:HS_crossing}. We'll return to this point below.

\paragraph{Angular dependence of divergence in the cross-channel OPE limit (for any fixed spin).}  
The behavior \eqref{eq:H_limit} of $\calH_{13|24}^{\text{single}(s)}$ in the OPE limit $\mathbf{r}_{12}\to 0$ differs from $\calA^{(s)}_{13|24}$ not only in its degree of divergence, but also in the angular dependence of the divergence's coefficient. For instance, while the divergent piece of $\calA^{(0)}_{13|24}$ only depends on the distances $r_{12},r_{13},r_{14}$, we see that $\calH_{13|24}^{\text{single}(0)}$ also depends on the direction of the small vector $\mathbf{r}_{12}$, via the argument of the $K$ function in \eqref{eq:H_limit}. In particular, when the direction of $\mathbf{r}_{12}$ is tuned so that the geodesics intersect, the $K$ function in $\calH_{13|24}^{\text{single}(0)}$ goes to infinity, while $\calA^{(0)}_{13|24}$ doesn't care in the least.

\paragraph{Behavior of the sum over spins.}  
We now come to what may be the most crucial difference: the spin-dependence of  $\calH_{13|24}^{\text{single}(s)}$ vs. $\calA^{(s)}_{13|24}$ in the $\mathbf{r}_{12}\to 0$ limit. Since the limiting expression \eqref{eq:H_limit} is spin-independent, its sum over spins diverges as $\sum_s 1$. This is consistent with $\calH_{13|24}^{\text{single}}$ diverging faster than the individual-spin contributions $\calH_{13|24}^{\text{single}(s)}$, i.e. as $\sim 1/r_{12}$ rather than $\sim 1/\sqrt{r_{12}}$. On the other hand, in section \ref{sec:Euclid:ell_0} we saw that the bound on $\calA_{13|24}^{(s)}$ decreases with spin as $\sim 4^{-s}$, making the sum over spins converge, so that $\calA_{13|24}$ has the same $O(\ln r_{12})$ behavior as the individual-spin contributions $\calA_{13|24}^{(s)}$. Unfortunately, we don't have a clear bulk intuition for this difference between $\calH_{13|24}^{\text{single}(s)}$ and $\calA_{13|24}^{(s)}$. However, we can identify its origin in the language of spectral $d\nu$ integrals, such as the one in \eqref{eq:exchange}. From eqs. (4.19),(5.19) of \cite{Bekaert:2015tva}, we see that the $d\nu$ integrands that produce $\calH_{ij|kl}^{\text{single}(s)}$ and $\calA_{ij|kl}^{(s)}$ are related as:
\begin{widetext}
\begin{align}
    \frac{d\nu\text{ integrand for }\calA_{ij|kl}^{(s)}}{d\nu\text{ integrand for }\calH_{ij|kl}^{\text{single}(s)}} 
      = \frac{\Gamma\!\left(\frac{1}{4}+\frac{s}{2}+\frac{i\nu}{2}\right)^2\Gamma\!\left(\frac{1}{4}+\frac{s}{2}-\frac{i\nu}{2}\right)^2}{\pi\,\Gamma(s)^2} \ . \label{eq:spectral_ratio}
\end{align}
\end{widetext}
At the single-trace pole $\nu=-i\big(s-\frac{1}{2}\big)$, the ratio \eqref{eq:spectral_ratio} evaluates to 1, as it should. However, as we saw in section \ref{sec:Euclid:ell_0}, when considering cross-channels, it is more helpful to consider the integral over real $\nu$ directly, rather than its pole structure. At real $\nu$ and large $s$, Stirling's formula gives:
\begin{align}
   \frac{d\nu\text{ integrand for }\calA_{ij|kl}^{(s)}}{d\nu\text{ integrand for }\calH_{ij|kl}^{\text{single}(s)}} \approx \frac{\left(e^{-2t\arctan t}(1+t^2)\right)^s}{4^{s-1}\sqrt{1+t^2}} \ , \nonumber
\end{align}
where we again denoted $t\equiv \nu/s$. Here, we can clearly see the $\sim 4^{-s}$ difference in scaling for the bulk exchanges $\calA_{ij|kl}^{(s)}$ vs. their single-trace parts $\calH_{ij|kl}^{\text{single}(s)}$ in the cross-channels.


\begin{thebibliography}{99}

\bibitem{Vasiliev:1990en}
M.~A.~Vasiliev,
``Consistent equation for interacting gauge fields of all spins in (3+1)-dimensions,''
Phys. Lett. B \textbf{243}, 378-382 (1990)
doi:10.1016/0370-2693(90)91400-6

\bibitem{Vasiliev:1995dn} 
M.~A.~Vasiliev,
``Higher spin gauge theories in four-dimensions, three-dimensions, and two-dimensions,''
Int.\ J.\ Mod.\ Phys.\ D {\bf 5}, 763 (1996)
[hep-th/9611024].

\bibitem{Vasiliev:1999ba} 
M.~A.~Vasiliev,
``Higher spin gauge theories: Star product and AdS space,''
In *Shifman, M.A. (ed.): The many faces of the superworld* 533-610
[hep-th/9910096].

\bibitem{Klebanov:2002ja} 
I.~R.~Klebanov and A.~M.~Polyakov,
``AdS dual of the critical O(N) vector model,''
Phys.\ Lett.\ B {\bf 550}, 213 (2002)
[hep-th/0210114].

\bibitem{Maldacena:1997re} 
J.~M.~Maldacena,
``The Large N limit of superconformal field theories and supergravity,''
Int.\ J.\ Theor.\ Phys.\  {\bf 38}, 1113 (1999)
[Adv.\ Theor.\ Math.\ Phys.\  {\bf 2}, 231 (1998)]
doi:10.1023/A:1026654312961
[hep-th/9711200].

\bibitem{Gubser:1998bc}
S.~S.~Gubser, I.~R.~Klebanov and A.~M.~Polyakov,
``Gauge theory correlators from noncritical string theory,''
Phys. Lett. B \textbf{428}, 105-114 (1998)
doi:10.1016/S0370-2693(98)00377-3
[arXiv:hep-th/9802109 [hep-th]].

\bibitem{Witten:1998qj} 
E.~Witten,
``Anti-de Sitter space and holography,''
Adv.\ Theor.\ Math.\ Phys.\  {\bf 2}, 253 (1998)
[hep-th/9802150].

\bibitem{Aharony:1999ti} 
O.~Aharony, S.~S.~Gubser, J.~M.~Maldacena, H.~Ooguri and Y.~Oz,
``Large N field theories, string theory and gravity,''
Phys.\ Rept.\  {\bf 323}, 183 (2000)
doi:10.1016/S0370-1573(99)00083-6
[hep-th/9905111].

\bibitem{Anninos:2011ui} 
D.~Anninos, T.~Hartman and A.~Strominger,
``Higher Spin Realization of the dS/CFT Correspondence,''
Class.\ Quant.\ Grav.\  {\bf 34}, no. 1, 015009 (2017)
doi:10.1088/1361-6382/34/1/015009
[arXiv:1108.5735 [hep-th]].

\bibitem{Sezgin:2002rt}
E.~Sezgin and P.~Sundell,
``Massless higher spins and holography,''
Nucl. Phys. B \textbf{644}, 303-370 (2002)
[erratum: Nucl. Phys. B \textbf{660}, 403-403 (2003)]
doi:10.1016/S0550-3213(02)00739-3
[arXiv:hep-th/0205131 [hep-th]].

\bibitem{Sezgin:2003pt}
E.~Sezgin and P.~Sundell,
``Holography in 4D (super) higher spin theories and a test via cubic scalar couplings,''
JHEP \textbf{07}, 044 (2005)
doi:10.1088/1126-6708/2005/07/044
[arXiv:hep-th/0305040 [hep-th]].

\bibitem{Aharony:2011jz}
O.~Aharony, G.~Gur-Ari and R.~Yacoby,
``d=3 Bosonic Vector Models Coupled to Chern-Simons Gauge Theories,''
JHEP \textbf{03}, 037 (2012)
doi:10.1007/JHEP03(2012)037
[arXiv:1110.4382 [hep-th]].

\bibitem{Giombi:2011kc}
S.~Giombi, S.~Minwalla, S.~Prakash, S.~P.~Trivedi, S.~R.~Wadia and X.~Yin,
``Chern-Simons Theory with Vector Fermion Matter,''
Eur. Phys. J. C \textbf{72}, 2112 (2012)
doi:10.1140/epjc/s10052-012-2112-0
[arXiv:1110.4386 [hep-th]].

\bibitem{Giombi:2012ms} 
S.~Giombi and X.~Yin,
``The Higher Spin/Vector Model Duality,''
J.\ Phys.\ A {\bf 46}, 214003 (2013)
doi:10.1088/1751-8113/46/21/214003
[arXiv:1208.4036 [hep-th]].

\bibitem{Fronsdal:1978rb}
C.~Fronsdal,
``Massless Fields with Integer Spin,''
Phys. Rev. D \textbf{18}, 3624 (1978)
doi:10.1103/PhysRevD.18.3624

\bibitem{Fronsdal:1978vb}
C.~Fronsdal,
``Singletons and Massless, Integral Spin Fields on de Sitter Space (Elementary Particles in a Curved Space. 7.,''
Phys. Rev. D \textbf{20}, 848-856 (1979)
doi:10.1103/PhysRevD.20.848

\bibitem{Fradkin:1986qy}
E.~S.~Fradkin and M.~A.~Vasiliev,
``Cubic Interaction in Extended Theories of Massless Higher Spin Fields,''
Nucl. Phys. B \textbf{291}, 141-171 (1987)
doi:10.1016/0550-3213(87)90469-X

\bibitem{Fradkin:1987ks}
E.~S.~Fradkin and M.~A.~Vasiliev,
``On the Gravitational Interaction of Massless Higher Spin Fields,''
Phys. Lett. B \textbf{189}, 89-95 (1987)
doi:10.1016/0370-2693(87)91275-5

\bibitem{Sleight:2016dba} 
C.~Sleight and M.~Taronna,
``Higher Spin Interactions from Conformal Field Theory: The Complete Cubic Couplings,''
Phys.\ Rev.\ Lett.\  {\bf 116}, no. 18, 181602 (2016)
doi:10.1103/PhysRevLett.116.181602
[arXiv:1603.00022 [hep-th]].

\bibitem{Bekaert:2015tva} 
X.~Bekaert, J.~Erdmenger, D.~Ponomarev and C.~Sleight,
``Quartic AdS Interactions in Higher-Spin Gravity from Conformal Field Theory,''
JHEP {\bf 1511}, 149 (2015)
doi:10.1007/JHEP11(2015)149
[arXiv:1508.04292 [hep-th]].

\bibitem{Sleight:2017pcz} 
C.~Sleight and M.~Taronna,
``Higher-Spin Gauge Theories and Bulk Locality,''
Phys.\ Rev.\ Lett.\  {\bf 121}, no. 17, 171604 (2018)
doi:10.1103/PhysRevLett.121.171604
[arXiv:1704.07859 [hep-th]].

\bibitem{Fitzpatrick:2012yx}
A.~L.~Fitzpatrick, J.~Kaplan, D.~Poland and D.~Simmons-Duffin,
``The Analytic Bootstrap and AdS Superhorizon Locality,''
JHEP \textbf{12}, 004 (2013)
doi:10.1007/JHEP12(2013)004
[arXiv:1212.3616 [hep-th]].

\bibitem{Fitzpatrick:2014vua}
A.~L.~Fitzpatrick, J.~Kaplan and M.~T.~Walters,
``Universality of Long-Distance AdS Physics from the CFT Bootstrap,''
JHEP \textbf{08}, 145 (2014)
doi:10.1007/JHEP08(2014)145
[arXiv:1403.6829 [hep-th]].

\bibitem{Alday:2010zy}
L.~F.~Alday, B.~Eden, G.~P.~Korchemsky, J.~Maldacena and E.~Sokatchev,
``From correlation functions to Wilson loops,''
JHEP \textbf{09}, 123 (2011)
doi:10.1007/JHEP09(2011)123
[arXiv:1007.3243 [hep-th]].

\bibitem{Alday:2015ota}
L.~F.~Alday and A.~Zhiboedov,
``Conformal Bootstrap With Slightly Broken Higher Spin Symmetry,''
JHEP \textbf{06}, 091 (2016)
doi:10.1007/JHEP06(2016)091
[arXiv:1506.04659 [hep-th]].

\bibitem{Maldacena:2002vr} 
J.~M.~Maldacena,
``Non-Gaussian features of primordial fluctuations in single field inflationary models,''
JHEP {\bf 0305}, 013 (2003)
doi:10.1088/1126-6708/2003/05/013
[astro-ph/0210603]

\bibitem{Bekaert:2014cea}
X.~Bekaert, J.~Erdmenger, D.~Ponomarev and C.~Sleight,
``Towards holographic higher-spin interactions: Four-point functions and higher-spin exchange,''
JHEP \textbf{03}, 170 (2015)
doi:10.1007/JHEP03(2015)170
[arXiv:1412.0016 [hep-th]].

\bibitem{Heemskerk:2009pn}
I.~Heemskerk, J.~Penedones, J.~Polchinski and J.~Sully,
``Holography from Conformal Field Theory,''
JHEP \textbf{10}, 079 (2009)
doi:10.1088/1126-6708/2009/10/079
[arXiv:0907.0151 [hep-th]].

\bibitem{Heemskerk:2010ty}
I.~Heemskerk and J.~Sully,
``More Holography from Conformal Field Theory,''
JHEP \textbf{09}, 099 (2010)
doi:10.1007/JHEP09(2010)099
[arXiv:1006.0976 [hep-th]].

\bibitem{El-Showk:2011yvt}
S.~El-Showk and K.~Papadodimas,
``Emergent Spacetime and Holographic CFTs,''
JHEP \textbf{10}, 106 (2012)
doi:10.1007/JHEP10(2012)106
[arXiv:1101.4163 [hep-th]].

\bibitem{Leonhardt:2003qu}
T.~Leonhardt, R.~Manvelyan and W.~Ruhl,
``The Group approach to AdS space propagators,''
Nucl. Phys. B \textbf{667}, 413-434 (2003)
doi:10.1016/j.nuclphysb.2003.07.007
[arXiv:hep-th/0305235 [hep-th]].

\bibitem{Leonhardt:2003sn}
T.~Leonhardt, W.~Ruhl and R.~Manvelyan,
``The Group approach to AdS space propagators: A Fast algorithm,''
J. Phys. A \textbf{37}, 7051 (2004)
doi:10.1088/0305-4470/37/27/013
[arXiv:hep-th/0310063 [hep-th]].

\bibitem{Costa:2014kfa}
M.~S.~Costa, V.~Gon\c{c}alves and J.~Penedones,
``Spinning AdS Propagators,''
JHEP \textbf{09}, 064 (2014)
doi:10.1007/JHEP09(2014)064
[arXiv:1404.5625 [hep-th]].

\bibitem{Hijano:2015zsa}
E.~Hijano, P.~Kraus, E.~Perlmutter and R.~Snively,
``Witten Diagrams Revisited: The AdS Geometry of Conformal Blocks,''
JHEP \textbf{01}, 146 (2016)
doi:10.1007/JHEP01(2016)146
[arXiv:1508.00501 [hep-th]].

\bibitem{Dyer:2017zef}
E.~Dyer, D.~Z.~Freedman and J.~Sully,
``Spinning Geodesic Witten Diagrams,''
JHEP \textbf{11}, 060 (2017)
doi:10.1007/JHEP11(2017)060
[arXiv:1702.06139 [hep-th]].

\bibitem{David:2020fea}
A.~David and Y.~Neiman,
``Bulk interactions and boundary dual of higher-spin-charged particles,''
JHEP \textbf{03}, 264 (2021)
doi:10.1007/JHEP03(2021)264
[arXiv:2009.02893 [hep-th]].

\bibitem{Lysov:2022zlw}
V.~Lysov and Y.~Neiman,
``Higher-spin gravity's ''string'': new gauge and proof of holographic duality for the linearized Didenko-Vasiliev solution,''
Accepted for publication in JHEP
[arXiv:2207.07507 [hep-th]].

\bibitem{Didenko:2008va}
V.~Didenko, A.~Matveev and M.~Vasiliev,
``Unfolded Description of AdS(4) Kerr Black Hole,''
Phys. Lett. B \textbf{665}, 284-293 (2008)
doi:10.1016/j.physletb.2008.05.067
[arXiv:0801.2213 [gr-qc]].

\bibitem{Didenko:2009td}
V.~Didenko and M.~Vasiliev,
``Static BPS black hole in 4d higher-spin gauge theory,''
Phys. Lett. B \textbf{682}, 305-315 (2009)
doi:10.1016/j.physletb.2009.11.023
[arXiv:0906.3898 [hep-th]].

\end{thebibliography}
\end{document}